\documentclass[twocolumn,showpacs,preprintnumbers,superscriptaddress,amsmath,amssymb,jcp]{revtex4}

\usepackage[pdftitle={appendix}, pdfauthor={Tobias Schmidt}, bookmarks=true]{hyperref}

\usepackage{graphicx}% Include figure files
\usepackage{dcolumn}% Align table columns on decimal point
\usepackage{bm}% bold math
\usepackage{color}
\usepackage[usenames,dvipsnames]{xcolor}
\usepackage{paralist}
\usepackage{verbatim}
\newcommand{\eps}[0]{\varepsilon}
\newcommand{\pphi}[0]{\varphi}
\newcommand{\up}[0]{\uparrow}
\newcommand{\dn}[0]{\downarrow}
\newcommand{\lp}[0]{\left}
\newcommand{\rp}[0]{\right}
\newcommand{\s}[0]{\sigma}

\newcommand{\LSDA}[0]{{L\!S\!D\!A}}

\newcommand{\rarr}[0]{\rightarrow}
\newcommand{\rr}[0]{(\mathbf{r})}
\newcommand{\rrp}[0]{(\mathbf{r}')}
\newcommand{\darsec}[0]{\texttt{DARSEC}}
\newcommand{\turbomole}[0]{\texttt{Turbomole}}

\newcommand{\rb}[0]{\mathbf{r}}

\begin{document}

\title{A self-interaction-free local hybrid functional: accurate binding energies \\ vis-\`a-vis accurate ionization potentials from Kohn-Sham eigenvalues}

\author{Tobias Schmidt}
\thanks{These authors contributed equally}
\affiliation{Theoretical Physics IV, University of Bayreuth, 95440 Bayreuth, Germany}

\author{Eli Kraisler}
\thanks{These authors contributed equally}
\affiliation{Department of Materials and Interfaces, Weizmann Institute of Science, Rehovoth 76100,Israel}

\author{Adi Makmal}
\altaffiliation[Present address: ]{Institut f\"ur Theoretische Physik, Universit\"at Innsbruck, Technikerstra{\ss}e 25,
 A-6020 Innsbruck, Austria}
\affiliation{Department of Materials and Interfaces, Weizmann Institute of Science, Rehovoth 76100,Israel}

\author{Leeor Kronik}
\affiliation{Department of Materials and Interfaces, Weizmann Institute of Science, Rehovoth 76100,Israel}

\author{Stephan K\"{u}mmel}
\affiliation{Theoretical Physics IV, University of Bayreuth, 95440 Bayreuth, Germany}

\date{\today}% It is always \today, today, but any date may be explicitly specified

\begin{abstract}
We present and test a new approximation for the exchange-correlation (xc) energy of Kohn-Sham density functional theory. It combines exact exchange with a compatible non-local correlation functional. The functional is by construction free of one-electron self-interaction, respects constraints derived from uniform coordinate scaling, and has the correct asymptotic behavior of the xc energy density. 
It contains one parameter that is not determined \emph{ab initio}. We investigate whether it is possible to construct a functional that yields accurate binding energies and affords other advantages, specifically Kohn-Sham eigenvalues that reliably reflect ionization potentials.
Tests for a set of atoms and small molecules show that within our local-hybrid form accurate binding energies can be achieved by proper optimization of the free parameter in our functional, along with an improvement in dissociation energy curves and in Kohn-Sham eigenvalues. However, the correspondence of the latter to experimental ionization potentials is not yet satisfactory, and if we choose to optimize their prediction, a rather different value of the functional's parameter is obtained. We put this finding in a larger context by discussing similar observations for other functionals and possible directions for further functional development that our findings suggest.

\end{abstract}

% PACS, the Physics and Astronomy Classification Scheme.
\maketitle

\section{Introduction}\label{sec.intro} 

Kohn-Sham (KS) density-functional theory (DFT)\cite{Hohenberg1964,Kohn1965} has become one of the most frequently used theories for electronic structure calculations. It employs the electron ground-state density, $n\rr$, as the central quantity and accounts for
all electronic interaction beyond 
the classical electrostatic (Hartree) repulsion, $E_\mathrm{H}$, via
the exchange-correlation (xc) energy functional, $E_\mathrm{xc}[n]$ \cite{Parr1989,Dreizler1995,Primer}.
Even though the xc energy is typically the smallest component in the ground-state total energy, it governs
binding properties, geometrical structures, and ionization processes \cite{Kurth2000,Burke2012,Primer}.
Thus, the quality of a DFT calculation depends decisively on the functional approximation put to task.

It has become popular to categorize density functional approximations according to the ``Jacob's ladder'' scheme introduced in Ref.~\cite{PerdewSchmidt2001}. 
Typically, the accuracy of a density functional approximation (DFA) improves when more ``ingredients'' are allowed in the functional construction, at the price of increased complexity. 
The local spin-density approximation (LDA)~\cite{Kohn1965}, which approximates $E_{xc}[n]$ based on the xc energy of the homogeneous electron gas \cite{Ceperley1980,PW'92,PZ'81,Vosko1980}, and even more so the 
semi-local generalized gradient approximations (GGAs)~\cite{PW1986,Becke1988,LYP88,PW1991,Perdew1996,Wu2006,Haas2011}, 
which additionally take the density gradient into account, offer a favorable ratio of computational expense and accuracy~\cite{Burke2012}. Hybrid functionals~\cite{Becke1993,Becke1993a,Perdew1996b,B3LYP_2,Adamo1999a,ErnScu} 
typically reach yet greater accuracy by combining a fixed percentage of Fock exchange
\begin{equation}
E_x^{ex}=-\frac{1}{2}\sum_{\substack{ i,j=1\\
\sigma=\up,\dn}}^{N_{\sigma}}\int\!\!\!\int
\frac{\varphi_{i\sigma}^*\rr\varphi_{j\sigma}\rr\varphi_{i\sigma}  \rrp\varphi_{
j\sigma}^*\rrp}{|\mathbf{r}-\mathbf{r}'|}\,\mathrm d^3r\mathrm d^3r',
\label{exx}
\end{equation}
with (semi-)local exchange and correlation energy terms (Hartree atomic units are used throughout).
Eq.\ (\ref{exx}) evaluated with the exact Kohn-Sham orbitals defines the exact Kohn-Sham exchange energy. Self-consistent Kohn-Sham calculations based on the energy of Eq.\ (\ref{exx}) use the optimized effective potential (OEP) equation (see \cite{Grabo1997,EngelDreizler2011,Kummel2008} and references therein). When we use the abbreviation EXX in the following, we always refer to this Kohn-Sham variant of exact exchange.

While the aforementioned functionals in many cases predict binding energies and bond-lengths reliably, semi-local DFAs and to some extent also hybrid functionals are less reliable for ionization processes, photoemission spectra and densities of states. Very early on it was realized that this problem is closely related to the (one-electron) self-interaction (SI)~\cite{PZ'81} error, i.e., to the fact that in exact 
DFT $E_\mathrm{xc}+E_\mathrm{H}$ should vanish for any one-electron system, but does not do so for these DFAs. Due to the SI error and the fact that semi-local functionals ``average over'' the derivative discontinuity \cite{levy83,averageDD}, the Kohn-Sham eigenvalues of the above mentioned approximate functionals typically fulfill neither the exact condition that the highest occupied eigenvalue $\varepsilon_\mathrm{ho}$ should match the first ionization potential (IP)~\cite{PPLB82,Levy1984,Almbladh1985,PerdewLevy97}, nor the approximate but for practical purposes equally important condition that upper valence eigenvalues are good approximations to higher IPs when they are calculated from accurate xc potentials \cite{Stowasser1999,chong,PRBRC09,PRBRC09E,dauthprl,bleiziffer}.

Although the interpretation of occupied eigenvalues even with the exact xc potential is approximate (except for $\varepsilon_\mathrm{ho}$), it is of great practical importance. For example, the band-structure interpretation of Kohn-Sham eigenvalues has had a great impact on solid-state physics and materials science~\cite{CohenChel88}. In recent years the interpretation of eigenvalues has become particularly important in the field of molecular semiconductors and organic electronics. Efforts to understand, e.g., photoemission experiments, have revealed severe shortcomings of traditional DFAs that go considerably beyond a spurious global shift of the eigenvalue spectrum~\cite{dori06,pesoep,mundttdpes,PRBRC09,PRBRC09E,ksvgks,dauthprl,MaromJCP2009,MaromAPA2009a,MaromJCTC2010,
Bisti2011}. A similar problem is witnessed also in solid state systems~\cite{Rinke2005,Fuchs2007,Fuchs2008,Rodl2009,Betzinger2012,Betzinger2013}. 
We emphasize that these problems of interpretation arise already for the occupied eigenvalues, i.e., the issues are separate from the well known band-gap problem~\cite{levy83,Sham1983,Kummel2008,Kronik2012} of Kohn-Sham theory.
The KS EXX potential leads to band structures and eigenvalues that match experiments much better than the eigenvalues from (semi-)local approximations~\cite{Bylander1995,Bylander1996,Grabo1997,gorlingEXXgaps,pesoep,engelEXXbands}. 

A comparison to hybrid functionals is more involved, because already the occupied eigenvalue spectrum depends sensitively on whether one uses the hybrid functional in a KS or a generalized KS calculation \cite{ksvgks}. For well understood reasons \cite{levy83,Sham1983}, the differences between the KS and the generalized KS eigenvalues become yet larger for unoccupied eigenvalues (see, e.g., the review \cite{Kronik2012}).This article's focus is on Kohn-Sham theory, therefore we do not discuss the comparison to hybrid functionals used in the generalized KS scheme in detail. We note, however, that in particular range-separated hybrid functionals used in the generalized KS approach can predict gaps and band structures quite accurately, as discussed, e.g., in \cite{hsegaps,Kronik2012}, but global hybrid functionals tend to yield a less reliable density of states for complex systems than self-interaction free Kohn-Sham potentials \cite{ksvgks,dauthprl}.

Besides these practical benefits, EXX also appears as 
a natural component of DFAs because it may be considered attractive to treat as many energy components as possible exactly, and including EXX has shown to be beneficial for, e.g., describing ionization, dissociation and charge transfer processes ~\cite{Kummel2008}.
However, bare EXX is a very poor approximation for binding energies, (see, e.g., Refs.~\cite{Engel2000d},\cite{EngelDreizler2011}, and \cite{Primer}, chapter 2). Combining EXX with a (semi-)local correlation term in many situations leads to results of inferior quality compared to pure EXX or semi-local DFAs, because of an imbalance between the delocalized exchange hole and the localized correlation hole~\cite{Gunnarsson1976,PerdewSchmidt2001,Prodan2005,Kummel2008}.

One promising approach for combining EXX with appropriate correlation in a balanced way is the local hybrid form
 \cite{Cruz1998,Jaramillo2003,Perdew2008}
\begin{equation}\label{eq.exc_lh}
e_{xc}(\rb)= (1-f[n](\rb)) e^{ex}_{x}(\rb) + f[n](\rb) e^{sl}_{x}(\rb) + e^{sl}_{c}(\rb).
\end{equation}
Here, $e_{xc}\rr$ is the xc energy density per particle that yields the xc energy via $E_{xc}[n]=\int n\rr\,e_{xc}\rr\,\mathrm d^3r$. The quantities $e^{sl}_{x}(\rb)$ and $e^{sl}_{c}(\rb)$ denote exchange and correlation energy densities per particle, respectively, approximated with (semi-)local expressions, 
whereas $e^{ex}_{x}(\rb)$ represents the EXX energy density per particle deduced from
Eq.~(\ref{exx}). The function $f[n]\rr$ is the local mixing function (LMF). It is a functional of the density and a decisive part of the local hybrid concept.

Eq.\ (\ref{eq.exc_lh}) can be viewed as a generalization of the common (global) hybrids. Instead of a fixed amount of EXX, the local hybrid can describe different spatial regions of a system with varying combinations of EXX and semi-local xc, by means of $f[n]\rr$ (where $0 \le f \le 1$). For example, whereas one-electron regions are supposed to be well-described using EXX, regions of slowly varying density are expected to be captured appropriately by (semi-)local xc functionals. 
The idea of local hybrids can also be understood in terms of the adiabatic connection theorem \cite{couplingconst}, because $f[n]\rr$ may offer further flexibility in an accurate construction of the 
coupling-constant-dependent xc energy~\cite{Cruz1998}, especially for small coupling constant values.

The local hybrid form was pioneered by Jaramillo \textsl{et al.}~\cite{Jaramillo2003} with a focus on reducing the one-electron SI-error in single-orbital regions. 
Numerous further local hybrid constructions followed~\cite{Arbuznikov2006,Arbuznikov2007,Bahmann07,Kaupp07, Perdew2008,ArbuzBahmKaupp09,haunschild2009,haunschild2010,haunschild2010_2,Theilacker11}. They proposed various LMFs with different one-electron-region indicators, suggested several (semi-)local exchange and correlation functionals to be used in the construction, and followed different procedures to satisfy known constraints and determine remaining free parameters.

In the present manuscript we propose a new local-hybrid approximation that combines full exact exchange with a compatible correlation functional. The development is guided by the philosophy of fulfilling known constraints \cite{Perdew2005a}: Our xc energy density per particle, $e_{xc}$, is one-electron SI-free, possesses the correct behaviour under uniform coordinate scaling, and has the right asymptotic behaviour at large distances. 
It includes one free parameter that is not determined uniquely from these constraints.

In difference to earlier work, our emphasis is not on improving further the accuracy of binding energies
 beyond the one that was achieved with global hybrids. Instead, we focus on whether it is possible to construct an approximation that yields binding energies of at least the same quality as established hybrids and at the same time affords other advantages, notably KS eigenvalues that approximate IPs reasonably well.
We find that if we choose the parameter in our functional by optimizing the prediction of binding energies, the latter are obtained with an accuracy that is similar to the one reached with usual global hybrids.
At the same time, we achieve a significant improvement in prediction of dissociation energy curves for selected systems. Improvement in prediction of the ionization energy via the highest occupied KS eigenvalue is also observed. It is especially large for alkali atoms. However, the quality of the ionization energy prediction is not yet satisfactory, and if we aim to optimize the prediction of the latter, a rather different value for the functional's free parameter is obtained. 
We put this finding in a larger context by discussing similar observations for other functionals.

The paper is organized as follows: Section~\ref{sec.theory} is devoted to the description
of the new local hybrid functional. Section~\ref{sec.numeric} (and the Appendix) provide methodological and computational details. Sec.~\ref{sec.results} presents and discusses the results, and Sec.~\ref{sec.conclusions} offers conclusions and a summary.

\section{Construction of the functional}\label{sec.theory} 

In the construction of our functional, we choose to concentrate on satisfying the following exact properties:
(i) use the concept of full exact exchange, as defined by the correct uniform coordinate scaling~\cite{Levy1985,Levy1991} (see elaboration below); 
(ii)   freedom from one-electron self-interaction \cite{PZ'81};
(iii)  correct asymptotic behavior of the xc energy density per particle at $|\mathbf{r}|\rarr\infty$~\cite{RvLBaerd94};
(iv)  reproduction of the homogeneous electron gas limit. 
In addition, we wish to maintain an overall balanced non-locality of exchange and correlation~\cite{Perdew2008}. 

Regarding property (i), under the uniform coordinate scaling $\rb \rarr \gamma \rb$ the density transforms as $n_\gamma(\rb) = \gamma^3 n(\gamma \rb)$, with its integral, $N$, unchanged and the exchange scales as $E_x[n_\gamma(\rb)] = \gamma E_x[n(\gamma \rb)]$~\cite{Primer,Kummel2008}, which implies $e_x^{ex}[n_\gamma(\rb)] = \gamma e^{ex}_x[n(\gamma \rb)]$. This scaling relation is fulfilled, e.g., by $e_x^\LSDA$, the exchange energy density per particle in the local spin-density approximation (LSDA). 

For the correlation functional, $E_c[n]$, no such simple scaling rule exists: the correlation scales as $E_c[n_\gamma(\rb)] = \gamma^2 E_c^{(1/\gamma)}[n(\rb)]$, where the superscript $(1/\gamma)$ indicates a system with an electron-electron interaction 
that is reduced by a factor of $\gamma$ \cite{Primer}. Additional scaling results for the correlation energy can be found in, e.g., Refs.~\cite{Levy1985,Levy1991}.
Here, we concentrate on the limiting case of high electron densities, i.e.\ $\gamma \rarr \infty$, where the xc energy should be dominated  by $E_x[n]$, \cite{Levy1991}
\begin{equation}\label{eq.fexx}
\lim_{\gamma\rightarrow\infty} \frac{E_{xc}[n_\gamma]}{E^{ex}_{x}[n_\gamma]}=1.
\end{equation}
A functional is said to use full exact exchange if it obeys Eq.\ (\ref{eq.fexx}) \cite{Perdew2008}. 

With this definition in mind, we return to  Eq.~(\ref{eq.exc_lh}). Using $e_{xc}(\rb) = e^{ex}_{x}(\rb) + e_c(\rb)$, we obtain
\begin{equation}\label{eq.ec_cc}
e_{c}(\rb) = f[n](\rb) \lp( e^{sl}_{x}(\rb) - e^{ex}_{x}(\rb) \rp) + e^{sl}_{c}(\rb).
\end{equation}
We now see that when $f[n]$ scales in the high density limit as $\gamma^a$ with $a<0$,
then it is clear that the first term on the RHS of Eq.~(\ref{eq.ec_cc}) is a correlation contribution rather
than an exchange term \footnote{Note that there exists a stronger requirement on the correlation energy, namely
$\lim_{\gamma \rarr \infty} E_c[n_\gamma] > -\infty$ (see Ref.~\cite{Levy1991}, Eq.(12)), which here we do not strive to fulfill.}. Assuming $e^{sl}_{c}(\rb)$ scales as $\gamma^b$ with $b<1$, the functional  $e_{xc}(\rb)$ that fulfills this condition can therefore justly be viewed as a combination of EXX, 
namely, $e^{ex}_{x}(\rb)$, and a compatible correlation term, $e_c(\rb)$. 

The reduced density gradient \cite{Perdew1996}
\begin{equation}\label{eq.red_den_grad}
t^2(\rb) := \lp( \frac{\pi}{3} \rp)^{1/3} \frac{a_0}{16 \Phi^2(\zeta(\rb))} \frac{|\nabla n(\rb)|^2}{n^{7/3}(\rb)},
\end{equation}
where $a_0$ is the Bohr radius, $\Phi(\zeta(\rb)) = \frac{1}{2} \lp( (1+\zeta)^{2/3}+(1-\zeta)^{2/3}\rp)$ and $\zeta(\rb) = (n_\up(\rb) - n_\dn(\rb))/(n_\up(\rb) + n_\dn(\rb))$ is the spin polarization, is a natural ingredient to be used to construct a $f[n](\rb)$ that aims at enforcing the uniform coordinate scaling, because in the high density limit $t^2 \sim \gamma$. We make use of this quantity as described in detail below.

Property (ii) is reflected in the equation $E_H[n_{i\sigma}]+E_{xc}[n_{i\sigma}]=0$, where
$n_{i\sigma}\rr=|\varphi_{i\sigma}\rr|^2$ are one-spin-orbital densities, with $\varphi_{i\sigma}\rr$ denoting the $i$-th KS-orbital in the spin-channel $\sigma$. One can attempt to realize such a one-spin-orbital condition by detecting regions of space in which the density is dominated by just one spin-orbital and making sure that full exact exchange and zero correlation is used there. Previous works have discussed the use of iso-orbital indicators~\cite{beckeiso,dobsoniso,pkzb,isoorb} for similar tasks.
Here, we define a one-spin-orbital-region indicator by
\begin{equation}\label{eq.ind}
d(\rb) = \frac{\tau_W(\rb)}{\tau(\rb)} \zeta^2(\rb),
\end{equation}
where $\tau_W(\rb) = |\nabla n(\rb)|^2 / (8n(\rb))$ is the von Weizs\"{a}cker kinetic energy
density and $\tau(\rb) = \frac{1}{2} \sum_\s \sum_{i=1}^{N_\s} |\nabla \pphi_{i\s}(\rb)|^2$ is the Kohn-Sham kinetic energy density.

For one-spin-orbital densities of ground-state character, $d(\rb) \rarr 1$, because $\tau(\rb) \rarr \tau_W(\rb)$ and $\zeta^2(\rb) \rarr 1$. 
For regions with slowly varying density, however, $d(\rb) \rarr 0$ because $\tau_W(\rb)$ tends to zero, whereas $\tau(\rb)$ does not.
In contrast to expressions suggested in the past~\cite{Jaramillo2003}, Eq.~(\ref{eq.ind}) does not classify a region of two spatially identical orbitals with opposite spins as a one-orbital region. It also avoids introducing~\cite{Arbuznikov2006,Arbuznikov2007,Bahmann07,Kaupp07, Perdew2008,ArbuzBahmKaupp09,Theilacker11} any parameters in $d\rr$.

Despite our use of Eq.~(\ref{eq.ind}) and the frequent use of similar indicators in the past, we wish to point out two caveats before proceeding. First, it should be noted that formally there exists a difference between \emph{one-electron} and \emph{one-spin-orbital} regions. The former correspond to spatial regions in the interacting-electrons system where the probability density is such that one finds just one electron. The latter, however, correspond to spatial regions in the KS system dominated by a single KS spin-orbital~\cite{Kurth1999}. There is no guarantee that these two regions coincide, because, strictly speaking, the interacting system and the KS
system have only the total electron density in common. 

Our second caveat refers to the fact that orbital densities are typically not of ground-state character. Therefore the equivalence of  $\tau(\rb)$ and $\tau_W(\rb)$ is not guaranteed for these over all space. It is reached, however, in the energetically relevant asymptotic region.  
We further note that it has recently been pointed out \cite{tdsicjcp} that also the Perdew-Zunger SI correction \cite{PZ'81} may have problems because of orbital densities not being ground-state densities. This may indicate that the question of how to associate orbitals with electrons for the purposes of eliminating self-interaction is a fundamental one, affecting all of the presently used concepts for self-interaction correction that we know of.

With the aim of fulfilling conditions (i) to (iv) we propose the following approximate form for 
our EXX-compatible correlation energy density per particle, $e_{c}(\rb)$:
\begin{align}\label{eq.ec_ISO}
\nonumber e_{c}(\rb)= \frac{1 - \frac{\tau_W(\rb)}{\tau(\rb)} \zeta^2(\rb)}{1+ct^2(\rb)} \lp( e^{\LSDA}_{x}(\rb) - e^{ex}_{x}(\rb) \rp) + \\
          + \lp( 1 - \frac{\tau_W(\rb)}{\tau(\rb)} \zeta^2(\rb) \rp) e^{\LSDA}_{c}(\rb).
\end{align}
In other words, we approximate the LMF function of Eq.~(\ref{eq.ec_cc}) by
\begin{equation}\label{eq.LMF}
f[n](\rb) = \frac{1 - d(\rb)}{1+ct^2(\rb)} = \frac{1 - \frac{\tau_W(\rb)}{\tau(\rb)} \zeta^2(\rb)}{1+ct^2(\rb)},
\end{equation}
the semi-local exchange energy density per particle by its LSDA form~\cite{Parr1989} $e^{sl}_{x}(\rb)=e^{\LSDA}_{x}(\rb)$,
and the semi-local correlation energy density per particle by
\begin{equation}\label{eq.ec.sl}
e^{sl}_{c}(\rb)= \lp( 1 - \frac{\tau_W(\rb)}{\tau(\rb)} \zeta^2(\rb) \rp) e^{\LSDA}_{c}(\rb),
\end{equation}
which is the LSDA correlation energy density per particle, multiplied by $(1-d(\rb))$.

The proposed functional is one-electron SI-free, has the required asymptotic
behavior for $e_{xc}\rr$ at $|\rb| \rarr \infty$, behaves correctly under uniform coordinate scaling, and reduces to the LSDA for regions of slowly varying density.

One-electron self-interaction is addressed via $d(\rb)$. When $d(\rb)$ tends to $1$, $e_c(\rb)$ vanishes and the only remaining term is $e^{ex}_x\rr$, which then cancels the Hartree repulsion.
Note that the semi-local correlation part, which is the last term in Eq.~(\ref{eq.ec_ISO}), also vanishes for one-spin-orbital regions. This is assured by introducing the prefactor $(1-d(\rb))$ in front of $e_c^\LSDA$. Otherwise, for one-orbital regions one would get the undesired, unbalanced combination of EXX and local correlation.

The correct uniform scaling is achieved due to the denominator in $f[n](\rb)$, which scales as $\gamma$, and cancels the $\gamma$-dependence
of the exchange terms that multiply it.
In addition, $e_c^\LSDA$ scales as $-\ln(\gamma)$ (see Eq.~(10) in Ref.~\cite{PW'92}, Sec.~II of Ref.~\cite{Levy1991}),
which is slower than $\gamma$. Therefore, the limit in Eq.~(\ref{eq.fexx}) is satisfied.

For slowly varying densities, $f[n](\rb) \rarr 1$ and $\tau_W(\rb) \rarr 0$, which yields $e_{xc}(\rb) \rarr e^{\LSDA}_{x}(\rb) + e^{\LSDA}_{c}(\rb)$,
reproducing the LSDA limit as required.

Finally, note that the proposed $e_{xc}$ approaches the known exact limit at $|\rb| \rarr \infty$.
Since EXX already has the right asymptotic decay of $e^{ex}_x(\rb) \sim -1/(2r)$ \cite{RvLBaerd94}, it suffices to verify that
$e_c(\rb)$ of Eq.~(\ref{eq.ec_ISO}) decays faster. Indeed, 
because the orbitals asymptotically tend to $\pphi_{i\s} \sim e^{-\alpha_{i\s}r}$, where $\alpha_{i\s} = \sqrt{-2
\eps_{i\s}}$, the density is dominated by the highest occupied orbital, $\pphi_{ho}$, and tends to $n \sim
|\pphi_{ho}|^2 \sim e^{-2\alpha_{ho} r}$. Because asymptotically $\tau_W/\tau \approx 1$ and
$t^2 \sim e^{\frac{2}{3} \alpha_{ho} r}$, one finds $f \sim t^{-2} \sim e^{-\frac{2}{3} \alpha_{ho} r}$, which
makes $e_c(\rb)$ decay exponentially. Therefore, the correct asymptotic behavior at $|\rb| \rarr \infty$ is achieved.

There remains one important point to be discussed. In Eq.~(\ref{eq.ec_ISO}) we are left with one undetermined parameter, $c$. Unfortunately, we presently do not know of an \emph{ab initio} constraint that would allow us to fix this parameter uniquely, although we do not rule out the possibility that future work may achieve this. 
The value of $c$ affects the amount of EXX that is used in a calculation and is therefore expected to have an influence in practical applications. One can therefore argue that not
having $c$ determined from first principles is a disadvantage. However, with $c$ being a free parameter, the functional form contains some freedom which allows one to adjust it to specific many-electron systems. One can therefore argue that our yet undetermined $c$ is in line with the principle of reducing (but not eliminating) empiricism in DFT~\cite{Burke2012}. 

In this first study, the freedom of varying $c$ will be used deliberately to explore the properties of the proposed functional.
We perform fitting of $c$ per system for a representative test set to observe how much its optimal value varies between the different systems, and whether a global fitting procedure, i.e.\ fitting for all systems combined, is at all justified.
In particular, we wish to elucidate the question of whether good binding energies and good eigenvalues can be achieved with the suggested local hybrid functional form. 
As an aside we note that when $c$ is a fixed, system-independent parameter, the proposed functional is fully size-consistent and complications that are known to occur with system-specific adjustment procedures \cite{aksizecon} are avoided.

\section{Methods} \label{sec.numeric} 

The proposed functional was implemented and tested using the
program package \darsec,~\cite{Makmal2009,Makmala} an all-electron code, which
allows for electronic structure calculations of single atoms or diatomic
molecules on a real-space grid represented by prolate spheroidal coordinates. We therefore avoid possible uncertainties associated with the use of pseudopotentials or complicated basis sets in OEP calculations~\cite{Kummel2008} -- an advantage for accurate functional testing. 

\darsec \, allows the user to solve the KS equations self-consistently for density- as well
as orbital-dependent functionals (ODFs), for example the proposed functional. For ODFs, the xc
potential is constructed by using either the full optimized effective
potential formalism (OEP) \cite{Kummel2008,Grabo1997} via the \textsl{S-iteration}-method
\cite{Kummel2003,Kummel2003a} or, with reduced computational effort, by employing the Krieger-Li-Iafrate (KLI) approximation~\cite{Li1992a}.
We note that other ways of defining approximations to the OEP exist \cite{lhf,CEDA2,staroverovprl2013}. However, for pure exchange earlier works have shown that total energies and eigenvalues are obtained with very high accuracy in the KLI approximation \cite{Li1992a,Grabo1997,lhf}, and for our local hybrid we explicitly compare KLI results to full OEP results in Sec.~\ref{sec.results.kli-oep} and find very good agreement.

In \darsec \,, all computations were converged up to $0.001$ Ry in the total energy, $E_{tot}$, as well as in the highest occupied KS eigenvalue, $\eps_{ho}$, by appropriately choosing the parameters of the real-space
grid and by iterating the self-consistent DFT cycle. For full OEP calculations, applying the
S-iteration method to the KLI xc potential typically resulted in a reduction of the maximum value of the
S-function \cite{Kummel2003} by a factor of 100. The spin and the axial angular momentum of the systems were
taken as in experiment. Note that to this end, for some systems it was necessary to force the KS occupation
numbers.

Numerical stability of self-consistent computations using ODFs, 
in the KLI- or OEP-scheme, mainly depends on the numerical realization of the functional derivative
\begin{equation}
 u_{i\sigma}\rr= \frac{1}{\varphi^*_{i\sigma}\rr}\frac{\delta
E^{}_{xc}[\{\varphi_{j\tau}\}]}{\delta\varphi_{i\sigma}\rr}, \label{eq.def_u_func}
\end{equation}
which ``conveys'' the special character of the corresponding xc functional into the calculation of the xc
potential. Because our functional approach results in a rather complicated function
$u_{i\sigma}\rr$ (see Appendix \ref{sec.app.pot}, Eq.~(\ref{eq.uisocomplete})), careful analytical
restructuring was necessary in
order to avoid diverging and unstable calculations. In particular, an explicit division by the KS orbitals
or the electron density should be avoided, because their exponential decay~\cite{Kreibich} leads to
instabilities at outer grid points. A numerically stable $u_{i\sigma}\rr$ was gained by such considerations,
for example by replacing $\tau_W(\rb) = |\nabla n(\rb)|^2 / (8n(\rb))$  in Eq.~(\ref{eq.ec_ISO}) with
the equivalent expression $\tau_W\rr=\frac{1}{2}|\nabla n^{\frac{1}{2}}\rr|^2$, or, in case division by the
density cannot be avoided, by equally balancing density terms of the same power in numerator and
denominator (for details see Appendix \ref{sec.app.pot}).

All results using (semi-)local functionals (LSDA~\cite{PW'92}, PBE~\cite{Perdew1996}) or the B3LYP hybrid functional~\cite{B3LYP_2} (evaluated within the generalized KS scheme~\cite{Seidl1996}) were obtained with the \turbomole \, program package \cite{TURBOMOLE}, using the def2-QZVPP basis set.
The pure EXX calculations were performed in \darsec \, by employing the functional derivative $u_{i\sigma}\rr$ originating from Eq.~(\ref{exx}) (as derived in Appendix \ref{sec.app.pot}, Eq.~(\ref{eq.exx3})).

When evaluating a new functional, it is reasonable to concentrate on a class of relatively simple systems to keep computational costs low and to refrain from additional sources of error 
beyond the
xc approximation, e.g., searching for an optimal geometry in systems with many degrees of freedom. However, the systems should not be too simple, so as to pose a significant challenge for the proposed functional. The class of
systems has to be large enough, as success or failure for one particular system has very limited meaning. It should also be rich enough to try to represent other systems that are not included. 
Previous work \cite{Perdew1996} has shown that a limited set of well selected small molecules can allow for meaningful exploration of a functional's properties.
For these reasons we focus on a set of 18 light diatomic molecules: H$_2$, LiH, Li$_2$, LiF, BeH, BH, BO, BF, CH,
CN, CO, NH, N$_2$, NO, OH, O$_2$, FH, F$_2$, and their constituent atoms. The systems include single-,
double-, and triple-bond molecules as well as atoms (no bonding). 

\section{Results}\label{sec.results} 

\subsection{Comparison of KLI and OEP} \label{sec.results.kli-oep}

While good agreement between the KLI and OEP scheme has been demonstrated before for ground-state energy
calculations using EXX \cite{Li1993}, the accuracy of the KLI approximation needs to be checked 
anew when it is applied to a previously untested functional. 
Table~\ref{table.oepcomp1} provides this check for our functional. It
compares the total energy and the highest occupied KS eigenvalue as obtained with the OEP and the KLI approximation for different values of the parameter $c$ (cf.\ Eq.~(\ref{eq.ec_ISO})) for different systems, and lists the corresponding differences for EXX for comparison. 

Table~\ref{table.oepcomp1} shows that the requirement $E_{tot}^{OEP} \le E_{tot}^{KLI}$~\cite{,Kummel2008} is fulfilled 
independent of the value of $c$ employed. Unlike for the total energy,
there is no theorem stating that the highest occupied KS eigenvalue found in the OEP scheme must be below its
KLI counterpart. For example, for the C atom and the N$_2$ molecule, we observe the opposite. Furthermore, because the suggested local hybrid with $c=0$ for spin-unpolarized systems ($\zeta\rr=0 \, \forall \, \rb$) is exactly equivalent to the purely semi-local constituent functional, one would expect the KLI and OEP results to coincide.
This is indeed fulfilled within numerical accuracy.
A detailed listing of the total energies and eigenvalues of the highest occupied KS states obtained by the KLI approximation in comparison to full OEP can be found in Appendix~\ref{sec.app.oep_kli_comp}, Tables~\ref{table.app.oepcomp1} and ~\ref{table.app.oepcomp2}.

With increasing $c$, a larger amount of EXX is employed and the functional gains more non-local
character, leading to greater deviations between KLI and OEP results. Note that, within the considered $c$-range, the
deviations with our functional are consequently lower than those obtained for EXX. The last statement applies to both $E_{tot}$ and $|\eps_{ho}|$. Furthermore, in agreement with Ref.~\cite{EngelDreizler2011} (p. 255), we observe an increasing difference between KLI and OEP results with growing number of electrons in the system.

To summarize, using the KLI approximation for our functional is as justified as it is for pure EXX. This observation is in agreement
with the fact that EXX is the limiting case of the suggested functional for $c\rightarrow \infty$.
 \begin{table}[ht]
 \caption{Comparison of total energy, $E$, and highest occupied KS eigenvalue, $\eps_{ho}$, obtained with the suggested local hybrid functional and with pure EXX, within both the KLI and OEP schemes, as a function of $c$ ($\Delta_E=E^{KLI}-E^{OEP}$, $\Delta_\eps=\eps^{KLI}_{ho}-\eps^{OEP}_{ho}$). All values are in Hartree.} \label{table.oepcomp1}
\centering
\begin{tabular*}{\columnwidth}{@{\extracolsep{\stretch{1}}}*{6}{ll|ccc|c}@{}}\hline
  &  &\multicolumn{3}{c}{suggested functional}\vline
&\multicolumn{1}{c} {EXX}\\ 
system &  & $c=0$ & $c=0.5$ & $c=2.5$ &\\ \hline
C	&$\Delta_E$	& 0.0000	& 0.0002 	& 0.0003	&0.0004	\\
	&$\Delta_\eps$	&-0.0005	& 0.0001 	& 0.0003	&0.0007	\\
BH	&$\Delta_E$	& 0.0000	& 0.0002 	& 0.0005	&0.0006	\\
	&$\Delta_\eps$	& 0.0000	& 0.0003 	& 0.0004	&0.0010	\\
Li$_2$	&$\Delta_E$	& 0.0000	& 0.0001 	& 0.0002	&0.0002	\\
	&$\Delta_\eps$	& 0.0000	& 0.0002 	& 0.0005	&0.0006	\\
NH	&$\Delta_E$	& 0.0001	& 0.0005 	& 0.0008	&0.0011	\\
	&$\Delta_\eps$	& 0.0007	& 0.0013 	& 0.0025	&0.0055	\\
N$_2$	&$\Delta_E$	& 0.0000	& 0.0009 	& 0.0017	&0.0023	\\
	&$\Delta_\eps$	& 0.0000	&-0.0010	&-0.0019	&0.0018	\\ \hline
\end{tabular*}
 \end{table}

\subsection{Fitting the parameter $c$ for each system}\label{sec.results.fit.per.sys}

The proposed functional has one unknown parameter, $c$. We aim to define a global value for $c$, relying on fitting it 
such that for a group of selected systems, some predefined quantity 
is optimally predicted (possible choices are discussed in detail below). 
As a prerequisite, we obtain individual $c$-values by optimizing the parameter for each of the systems separately.
As a test for whether a global fitting procedure is meaningful, we verify that these individual $c$-values are 
clustered within a reasonable numerical range.

In the following, we present two ways to fit $c$. One possibility is fitting the dissociation energy: To find $c$ for the molecule AB, the total energies of the molecule and its constituent atoms have to be calculated, with the same $c$. Then, the dissociation energy $D(c) = E_A(c)+E_B(c)-E_{AB}(c)$ is fitted to its experimental value~\cite{HandChemPhys92}, $D^{exp}$, by varying $c$. Alternatively, one can compute the total energy of the system for various values of $c$ and fit it to the experimental total energy. The latter is obtained for atoms as $E^{exp}_{atom} = -\sum_i I_i^{exp}$ - the sum of all its experimental IPs, $I_i^{exp}$; For molecules as $E^{exp}_{AB} = E^{exp}_A + E^{exp}_B - D^{exp}$. Unless explicitly stated otherwise, here and throughout molecular properties are calculated at their experimental bond lengths~\cite{HandChemPhys92}.

Table~\ref{table.cD_cE} presents optimized $c$-values for various systems, 
obtained from both the $D$-fitting and the
$E$-fitting procedures. 
The numerical uncertainty reported for the $c$-values is due to the 1 mRy numerical accuracy in the total energy. The table confirms that the chosen numerical accuracy for the total energy is indeed sufficient. 
We note that in the $E$-fitting there is a tendency for $c$ to increase with the electron number, which reflects a larger contribution of exact exchange. We attribute this to the fact that the energy of the core electrons (which is less important in $D$-fitting) is more strongly dominated by exchange. For our purposes, the most important conclusion to be drawn from Table~\ref{table.cD_cE} is that for all systems examined in both approaches, optimal values for $c$ lie between 0 and 1, and are never larger than 2. This observation justifies our pursuit of a global value of $c$ .

\begin{table}
  \centering
  \begin{tabular*}{\columnwidth}{@{\extracolsep{\stretch{1}}}*{3}{lcc}@{}}
  \hline
  System & $c_D$             & $c_E$             \\
  \hline
  H$_2$  & 0.552 $\pm$ 0.002  & 0.537  $\pm$ 0.012  \\
  LiH    & 0.642 $\pm$ 0.005  & 0.556 $\pm$ 0.004 \\
  Li$_2$ & 1.50  $\pm$ 0.06   & 0.571 $\pm$ 0.002 \\
  LiF    & 0.141 $\pm$ 0.006  & 0.976 $\pm$ 0.003 \\
  BeH    & 0.746  $\pm$ 0.025 & 0.648 $\pm$ 0.004 \\
  BH     & 0.590  $\pm$ 0.010 & 0.685 $\pm$ 0.004 \\
  BO     & 0.288 $\pm$ 0.007  & 0.916 $\pm$ 0.002 \\
  BF     & 0.578  $\pm$ 0.027 & 0.943 $\pm$ 0.002 \\
  CH     & 0.672  $\pm$ 0.028 & 0.741 $\pm$ 0.003 \\
  CN     & 0.146  $\pm$ 0.005 & 0.908 $\pm$ 0.002 \\
  CO     & 0.283  $\pm$ 0.009 & 0.916 $\pm$ 0.002 \\
  NH     & 0.667  $\pm$ 0.027 & 0.811 $\pm$ 0.004 \\
  N$_2$  & 0.107 $\pm$ 0.009  & 0.908 $\pm$ 0.003 \\
  NO     & 0.329  $\pm$ 0.009 & 0.960 $\pm$ 0.002 \\
  OH     & 1.20  $\pm$ 0.07   & 0.942 $\pm$ 0.004 \\
  O$_2$  & 0.472 $\pm$ 0.009  & 1.004 $\pm$ 0.002 \\
  FH     & 0.075 $\pm$ 0.011  & 1.105 $\pm$ 0.004 \\
  F$_2$  & 0.356 $\pm$ 0.006  & 1.206 $\pm$ 0.003 \\
  \hline
  H      & ---               & any               \\
  Li     & ---               & 0.543 $\pm$ 0.005 \\
  Be     & ---               & 0.644 $\pm$ 0.005 \\
  B      & ---               & 0.698 $\pm$ 0.003 \\
  C      & ---               & 0.757 $\pm$ 0.002 \\
  N      & ---               & 0.848 $\pm$ 0.005 \\
  O      & ---               & 0.925 $\pm$ 0.004 \\
  F      & ---               & 1.067 $\pm$ 0.003 \\
  \hline
\end{tabular*}
  \caption{The parameter $c$ optimized for various systems, using the $D$- and $E$-fitting procedures.}\label{table.cD_cE}
\end{table}

\subsection{Determining a global value for the parameter $c$}\label{sec.results.global.fit}

Following the conclusion that the parameter $c$ can indeed be fitted, we performed a series of calculations,
obtaining the $c$-dependent average relative errors
\begin{equation}\label{eq.delta_A}
    \delta_A(c) = \sqrt{ \frac{1}{M} \sum_{m=1}^M \lp( \frac{A_m(c) - A^{exp}}{A^{exp}} \rp)^2}.
\end{equation}
Here, $A$ can refer to the dissociation energy, $D$, the total energy, $E$, or the ionization potential $I$ evaluated
via $I= -\eps_{ho}$, the IP-theorem for the exact functional. The index $m$ runs over all the systems calculated~\footnote{The quantities $\delta_E$ and $\delta_I$ were obtained relying on all the molecules and atoms in the reference set ($M=26$), while $\delta_D$ was obtained relying on the molecules only ($M=18$).}.

\begin{figure}
  \includegraphics[scale=0.45,trim = 5mm 0mm 0mm 0mm]{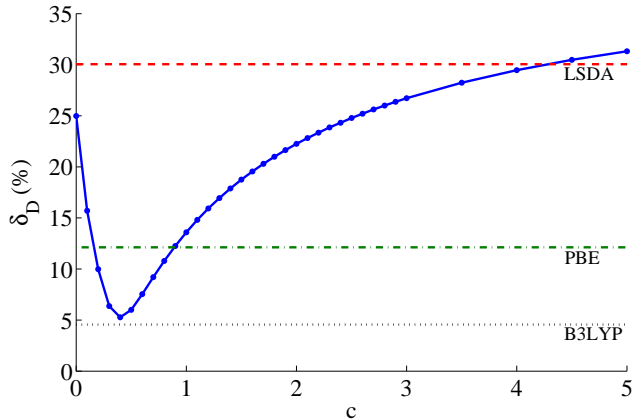}
  \caption{Average relative error of the dissociation energy, $\delta_D$, as a function of the parameter $c$ (solid line). Relative errors for the LSDA (dashed), PBE (dash-dotted) and B3LYP (dotted) functionals are given for comparison.
Pure EXX reaches an error of $\delta^{EXX}_D(c)=66\%$ and exceeds the scale we chose here.}\label{fig.deltaD}
\end{figure}
\begin{figure}
  \includegraphics[scale=0.45,trim = 5mm 0mm 0mm 0mm]{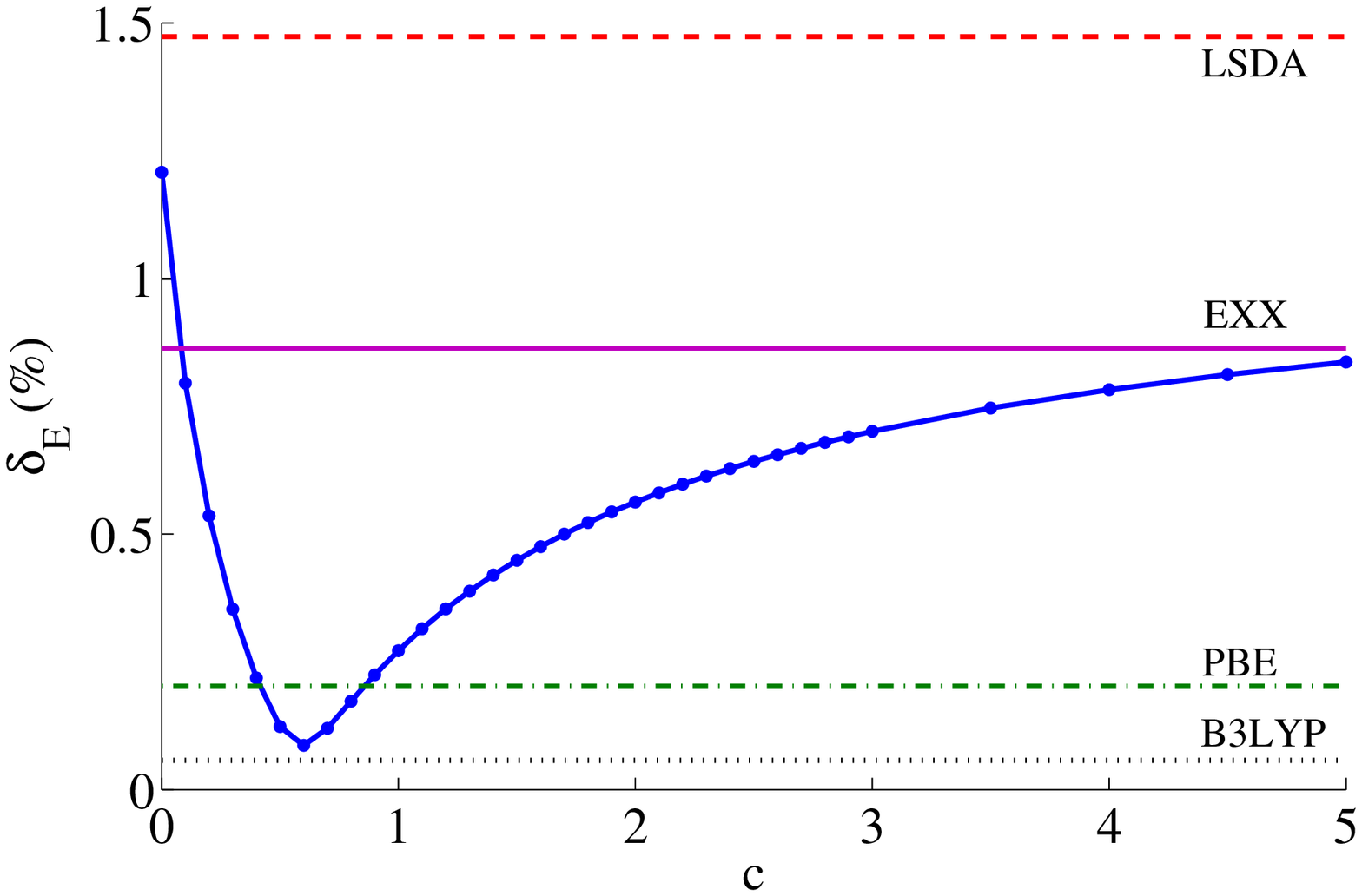}
  \caption{Average relative error of the total energy, $\delta_E$, as a function of the parameter $c$ (blue solid line). Relative errors for the LSDA (dashed), PBE (dash-dotted) and B3LYP (dotted) functionals, as well as pure EXX(KLI) (purple solid line) are given for comparison.}\label{fig.deltaE}
\end{figure}

The functions $\delta_D(c)$ and $\delta_E(c)$ are plotted in Figs.~\ref{fig.deltaD} and~\ref{fig.deltaE},
respectively, accompanied by the average relative errors for commonly used functionals: the LSDA, PBE, and B3LYP.
As mentioned previously, the B3LYP results here and in the following were obtained in the generalized KS approach, which we, based on previous experience \cite{Kummel2008}, expect to yield total energies that are very similar to the ones from the KS approach for the systems studied here.
For completeness, results obtained with pure EXX evaluated in the KLI approximation are also reported.

 In both figures we observe clear minima for the proposed functional at the values of $c_0 = 0.4$ for $\delta_D$ and $0.6$ for $\delta_E$, with minimal error values of $5.3\%$ and $0.09\%$, respectively.  These error values are close to those achieved with the B3LYP functional, and are significantly better than the PBE and LSDA results.
Because 
optimizing $\delta_D(c)$ and $\delta_E(c)$ demands almost the same value for $c$, a satisfying description of both properties is possible using a common parameter of $c=0.5$.
For this $c$, the relative error in the dissociation energy $\Delta D_m = (D_m - D^{exp})/D^{exp}$ is lowest for the BF molecule (0.7\%) and highest for Li$_2$  and F$_2$  (14\% and 17\%, respectively). The relative error in the total energy is more evenly spread around 0.12\%.

\begin{figure}
  \includegraphics[scale=0.45,trim = 5mm 0mm 0mm 0mm]{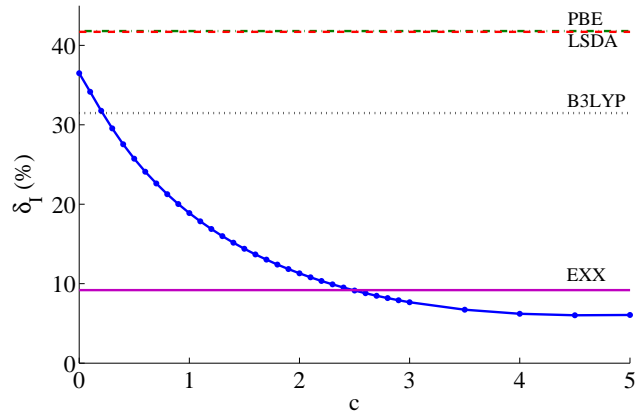}
  \caption{Average relative error of the IP predicted via the highest occupied KS eigenvalue, $\eps_{ho}$, as a function of the parameter $c$ (blue solid line). Relative errors for the LSDA (dashed), PBE (dash-dotted) and B3LYP (dotted) functionals, as well as pure EXX(KLI) (purple solid line) are given for comparison.}\label{fig.deltaEPS}
\end{figure}

The function $\delta_I(c)$ shown in Fig.~\ref{fig.deltaEPS} exhibits a different behavior, reaching its minimum of 6 \% at a higher value of $c\approx4.5$. 
\footnote{When calculating $\delta_I(c)$, the vertical experimental ionization potentials were used (see Ref.~\cite{HandChemPhys92} and \texttt{http://webbook.nist.gov})}
When evaluated at $c=0.5$, the average relative error is $\delta_I(c=0.5) = 26\%$.
Although lower than 31\% for B3LYP and 42\% for both LSDA and PBE, such a deviation is rather significant. Therefore, Fig.~\ref{fig.deltaEPS} suggests that in order to reach good agreement between the experimental IP and $-\eps_{ho}$, a larger amount of EXX is required.

Interestingly, when calculating NH and BO, we observed that the highest occupied state changes with varying the parameter $c$ from $\eps_{ho}=\eps_{3\downarrow}$ to $\eps_{5\uparrow}$ at approximately $c=0.7$ for NH, and from $\eps_{6\downarrow}$ to $\eps_{7\uparrow}$ at $c=1.6$ for the BO molecule. Such systems could therefore be good candidates for checking the functional's ability to predict physically meaningful orbitals in the sense of Ref.~\cite{dauthprl}.

Last, we checked the previously made assumption that experimental bond lengths can be used, assuming they are not very different from those obtained by relaxation. To this end, all 18 molecules in the reference set were relaxed, and the obtained bond lengths $L_m$ were compared to the experimental values, $L_m^{exp}$~\cite{HandChemPhys92}. 
It was found that for most systems $L_m^{exp}$ lies within the computational error for $L_m$ and the difference $|L_m - L_m^{exp}|$ is below 0.02 Bohr~\footnote{The numerical error in $L_m$ is governed by the accuracy of 1 mRy in the total energy, rather than by the convergence of the relaxation process.}, except F$_2$, 
where $|L_m - L_m^{exp}| \approx 0.08$ Bohr.
\footnote{Two exceptional cases are LiH and Li$_2$, which have an extremely shallow $E(L)$ minimum. The uncertainty of 1 mRy in the total energy translates into a numerical uncertainty in the bond length of 0.16 Bohr and 0.26 Bohr, respectively. Therefore, the difference $|L_m - L_m^{exp}|$, being 0.04 Bohr and 0.22 Bohr, although large, has no actual meaning due to the large numerical uncertainty.} 

\subsection{Achievements of the suggested functional}\label{sec.results.c0.5}

In the following, we examine some of the proposed functional's properties at the value of $c=0.5$, which was determined in Sec.\ref{sec.results.global.fit}. 

As the functional is one-electron SI-free (see Sec.~\ref{sec.theory}), it is important to investigate its
behavior in systems that are known to suffer from a large self-interaction error when described by standard DFAs. First, for one-electron systems, like H, He$^+$,
H$_2^+$, etc.\ the functional reduces analytically to the EXX functional, as can be seen from Eq.~(\ref{eq.ec_ISO}).
Therefore, all the properties of these systems are obtained, by construction, exactly. This advantage is not
shared by (semi-)local or most hybrid functionals.

\begin{figure}
  \includegraphics[scale=0.6,trim = 20mm 0mm 0mm 0mm]{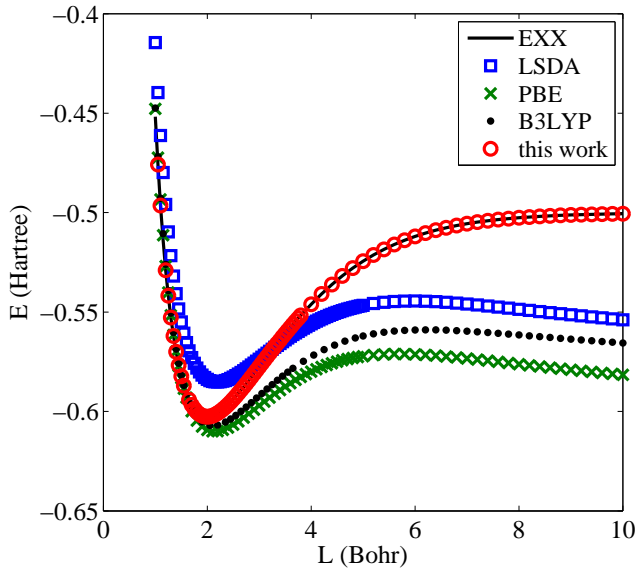}
  \caption{Dissociation curve of the H$_2^+$ molecule, for the LSDA (squares), PBE (x's), B3LYP (dots), EXX (solid line) and the suggested
functional (circles).}\label{fig.dissH2+}
\end{figure}

In particular, Fig.~\ref{fig.dissH2+} presents the dissociation curve of H$_2^+$, 
obtained with 
various functionals. It can be seen that, as expected, the curve obtained with the proposed functional perfectly agrees with the EXX curve, which provides the exact result in this case. In particular, our local hybrid does not exhibit a spurious maximum in the curve, which appears in 
conventional approximate 
functionals at bond lengths around 5-6 Bohr \cite{Bally1997,Ruzsinszky05,Cohen2008,Livshitz2008,Nafziger2013} and whose electrostatic origin has recently been discussed \cite{dwyer}. The dissociation of neutral H$_2$ is a special challenge for most density functionals and is closely connected to the question of how static correlation is accounted for~\cite{cohenchemrev}. Our local hybrid for H$_2$ yields a binding curve that is qualitatively similar to the one obtained in pure exchange calculations, i.e., for large internuclear separation the lowest energy is obtained with a spin-polarized atomic density centered around each nucleus, which yields a total energy of 1 Hartree. Quantitatively, there are differences with respect to the EXX solution: The point from which on the spin-polarized solution has a lower energy than the spin-unpolarized one lies at about 3.6 Bohr with our local hybrid, and the minimum energy is -1.173 Hartree as compared to -1.134 Hartree obtained with EXX.

Generally, delocalization in stretched molecular bonds is conceptually connected to the SI-error of DFAs~\cite{Cohen2008}. 
Therefore, reduction of this error marks a first step towards enhancing the description of dissociation processes and chemical reactions~\cite{Ruzsinszky2006}.

\begin{figure}
  \includegraphics[scale=0.6,trim = 10mm 0mm 0mm 0mm]{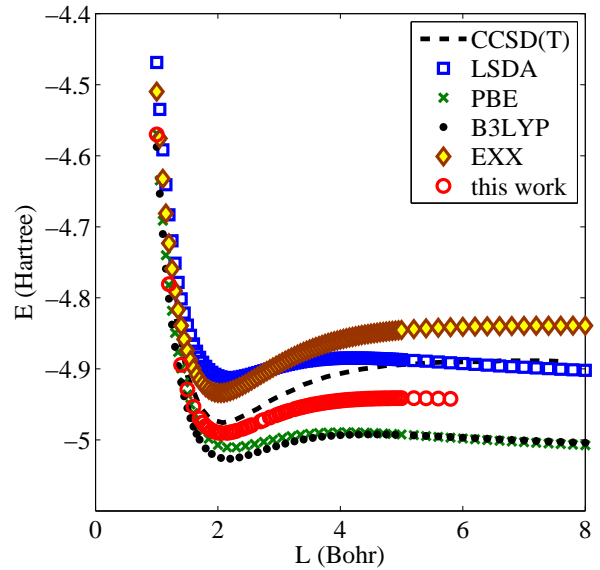}
  \caption{Dissociation curve of the He$_2^+$ molecule, for the LSDA (squares), PBE (x's), B3LYP (dots), EXX
(rhombi), and the suggested functional with $c=0.5$ (circles), compared to CCSD(T) results from Ref.~\cite{Ruzsinszky05} (dashed line)}\label{fig.dissHe2+}
\end{figure}

To examine this, the dissociation curve of the 3-electron molecule He$_2^+$ is shown in Fig.~\ref{fig.dissHe2+}.
The curve achieved with the suggested functional is the closest to the reference result
obtained with a highly accurate wavefunctions method (see Ref.~\cite{Ruzsinszky05} and references therein). Here
again, only our local hybrid and the EXX curves do not possess the spurious maximum, which appears in the 
conventional approximations -- LSDA, PBE and B3LYP around 4 Bohr. Unlike for the H$_2^+$ system, 
here the proposed functional does not
automatically reduce to the exact expression, and therefore the accurate prediction for He$_2^+$ in Fig.~\ref{fig.dissHe2+} can be seen as a consequence of the strong reduction of SI-errors, in agreement with a previous study.~\cite{Korzdorfer2008}

We further investigated how well  $\eps_{ho}$ corresponds to the experimental IP~\cite{HandChemPhys92} for the atoms Li, Na, and K. 
These atoms can be considered as quasi-one-electron systems,
consisting of electrons arranged in closed shells, which screen the charge of the nucleus, and one
additional electron in the last open shell. Table~\ref{table.closedshell} shows that the $\eps_{ho}$ obtained from our functional evaluated with $c=0.5$ is closer to the experimental IP than the  $\eps_{ho}$ from
 LSDA, PBE, and B3LYP. Note that for these systems a remarkable improvement is achieved, as one would expect
from their strong one-electron character.

 \begin{table}[ht]
 \caption{The highest occupied eigenvalue compared to the experimental IP for Li, Na, and K, computed with four different functionals (LSDA, PBE, B3LYP and our suggested functional using $c=0.5$). The table contains the absolute numbers in Hartree as well as the relative error in \%.}\label{table.closedshell}

\centering
\begin{tabular*}{\columnwidth}{@{\extracolsep{\stretch{1}}}*{6}{lc|rrrr}@{}}\hline
  & IP &\multicolumn{4}{c}{$-\eps_{ho}$}\\ 
system & Exp.  & LSDA & PBE & B3LYP& suggested functional\\ \hline
Li	&$0.1981$	& 0.1163 	& 0.1185	&0.1311 	& 0.1797\\
	&		&(-41 \%)	&(-40 \%)	&(-34 \%)	& (-9 \%)	\\
Na	&$0.1886$	& 0.1131 	& 0.1116	&0.1251 	& 0.1647\\
	&		&(-40 \%)	&(-41 \%)	&(-34 \%)	& (-13 \%)	\\
K 	&$0.1595$	& 0.0961 	& 0.0930	&0.1038 	& 0.1334\\
	&		&(-40 \%)	&(-42 \%)	&(-35 \%)	& (-16 \%)	\\\hline

\end{tabular*}
 \end{table}

\section{Conclusions}\label{sec.conclusions}

In this article we presented the construction of a local hybrid functional that combines full exact exchange with compatible correlation. The functional respects the homogeneous electron gas limit and addresses the one-electron self-interaction error via a one-spin-orbital-region indicator. The qualitative improvement that is achieved with this construction is reflected in, e.g., dissociation energy curves for H$_2^+$ and He$_2^+$ that are much more realistic than the ones obtained from  (semi-) local functionals and global hybrids. 
We investigated different conditions for fixing the undetermined parameter of the functional. When the parameter is fit to minimize binding energy errors or total energy errors, with respect to experiment, then our local hybrid reaches an accuracy that is better than LSDA or PBE and similar to the one afforded by the B3LYP global hybrid.
Predicting the first ionization energy via the highest occupied eigenvalue $\eps_{ho}$ is more accurate with our functional than with LSDA, PBE or B3LYP, but still not satisfactorily accurate.

When the parameter is fit such that $-\eps_{ho}$ should be as close as possible to the experimental first ionization potential, then the local hybrid functional achieves much smaller errors in this quantity than, e.g., B3LYP. However, the value obtained for the free parameter differs considerably from the one that was obtained by fitting to binding or total energies. As a result, prediction of these energies considerably differs from their experimental values. Therefore, our local hybrid does allow for reaching accurate binding energies or accurate highest eigenvalues, but not with the same functional parametrization.

Looking at this from a more general perspective, we note that many functionals can achieve good accuracy on one of the aforementioned properties or the other, but not on both properties at the same time. 

A first example are global hybrids. With the usual 0.2 to 0.25 fraction of exact exchange they yield good binding energies, but highest eigenvalues that are considerably too small in magnitude. 
Increasing the aforementioned fraction to $\sim 0.75$ leads to improved gap prediction~\cite{Sai2011,Imamura2011,Atalla2013}. 
However, such a large fraction of exact exchange can compromise significantly the accuracy in thermochemical~\cite{Adamo1999a,ErnScu,Zhao2008} or electronic structure properties.~\cite{MaromJCTC2010,ksvgks,Kronik2012}

A second example are range-separated hybrid functionals. When combined with a tuning procedure based on the IP theorem \cite{Stein2010,Refaely-Abramson2011,Refaely-Abramson2012,Kronik2012,Korzdorfer2011,Sini2011,Foster2012,Phillips2012, Risko2013}, 
they allow for obtaining eigenvalues that reflect ionization potentials very accurately by construction. However, the typical value of the range-separation parameter that is reached by such tuning is quite different from the one that is reached when atomization energy errors are minimized via the range-separation parameter \cite{LivshitzBaer2007}. 

A third example is provided by the various self-interaction correction schemes.
Different forms of self-interaction correction
greatly improve the interpretability of the eigenvalues when compared to semi-local functionals \cite{PZ'81,PRBRC09,pederson,Kluepfel2012,PGR,Korzdorfer2008,scuseria}, but binding energies are not
accurately predicted \cite{scuseria,hofmannkluepfel} unless the correction is ``scaled down'' \cite{scaleddown,Kluepfel2012}. 

This rather universal difficulty to achieve accurate eigenvalues and accurate binding energies at the same time may indicate that combining these two properties may require a type of physics that all present day functionals lack~\cite{Verma2012}.

Recent work provides two new and interesting perspectives on this problem. On the one hand, it has been noted that even a semi-local functional can yield eigenvalues that are qualitatively similar to the ones obtained from bare EXX when the asymptotic properties of a GGA are carefully determined \cite{newgga}. On the other hand, it has recently been shown that an ensemble perspective offers new and improved ways of interpreting eigenvalues and extracting information from semi-local functionals \cite{Kraisler2013}.
Exploring in particular this latter option, i.e., combining the 
ensemble approach with the present local hybrid functional, will be the topic of future work and may shed further light on the question of 
how to obtain accurate binding energies and Kohn-Sham eigenvalues from the same functional.

\begin{acknowledgments} 
S.K.\ gratefully acknowledges discussions with J.~P.~Perdew on local hybrids in general and on an early version of this functional in particular. We thank Baruch Feldman for fruitful discussions.
Financial support by the DFG Graduiertenkolleg 1640, the European Research Council, the German-Israeli Foundation, and the Lise Meitner center for computational chemistry is gratefully acknowledged. E.K.\ is a recipient of the Levzion scholarship. T.S.\ acknowledges support from the Elite Network of Bavaria (``Macromolecular Science'' program).

\end{acknowledgments}

\appendix

\section{Derivation of the correlation potential}\label{sec.app.pot} 
In order to employ the OEP formalism~\cite{Grabo1997,Kummel2008} (or its KLI approximation~\cite{Li1992a}), one has to provide an analytical expression for the functional derivative of the explicitly orbital-dependent exchange-correlation energy, $E_{xc}[ \{ \pphi_{i \s} \} ]$, with respect to the orbitals $\{ \pphi_{i \s} \}$. For the functional proposed in the 
present contribution,
\begin{equation}\label{func_exp_1}
 E_{xc} [ \{ \pphi_{i \s} \} ] = E_x^{ex}[ \{ \pphi_{i \s} \} ] + E^{iso}_{c} [ \{ \pphi_{i \s} \} ] + E_c^{sl} [ \{ \pphi_{i \s} \} ] ,
 \end{equation}
where $E_x^{ex}[ \{ \pphi_{i \s} \} ]$ is the exact exchange defined in Eq.(\ref{exx}), $E_c^{sl} [ \{ \pphi_{i \s} \} ]$ is the semi-local correlation energy, whose energy density per particle, $e_c^{sl}(\rb)$, is given in Eq.~(\ref{eq.ec.sl}), and $E^{iso}_{c} [ \{ \pphi_{i \s} \} ]$ equals
\begin{equation}\label{eq.Eiso}
E^{iso}_{c} [ \{ \pphi_{i \s} \} ] = \int f\rrp \, n\rrp \, \lp(e_x^\LSDA \rrp-e_x^{ex}\rrp\rp)\,\mathrm d^3r',
\end{equation} 
with the LMF function, $f(\rb)$, being defined in Eq.~(\ref{eq.LMF}).

Due to the additive structure of Eq.~(\ref{func_exp_1}), the functional derivative
\begin{equation}
u_{i\sigma}\rr = \frac{1}{\varphi^*_{i\sigma}\rr}\frac{\delta E_{xc}[ \{ \pphi_{i \s} \} ]}{\delta \varphi_{i\sigma}\rr} 
\end{equation}
can be split 
in three terms:
\begin{eqnarray}\label{allparts}
u_{i\sigma}\rr = u^{exx}_{i\sigma}\rr + u^{iso}_{i\sigma}\rr + u^{sl}_{i\sigma}\rr,
\end{eqnarray}
which are considered separately in the following.

\subsection{The exact exchange contribution}
The first term on the RHS of Eq.~(\ref{allparts}) can be computed directly from the exact-exchange expression (Eq.~(\ref{exx}))
\begin{equation}
E_x^{ex}=-\frac{1}{2}\sum_{\substack{
i,j=1\\
\sigma=\up,\dn}}^{N_{\sigma}}\int\int
\frac{\varphi_{i\sigma}^*\rr\varphi_{j\sigma}\rr\varphi_{i\sigma}  	\rrp\varphi_{
j\sigma}^*\rrp}{|\mathbf{r}-\mathbf{r}'|}\,\mathrm d^3r\mathrm d^3r'
\label{exx2}
\end{equation}
and simply reads
\begin{equation}
\varphi^*_{i\sigma}\rr\,u^{exx}_{i\sigma}\rr =
-\sum_{j=1}^{N_{\sigma}}\varphi_{j\sigma}^*\rr\int
\frac{\varphi^*_{i\sigma}\rrp\varphi_{j\sigma}\rrp}{|\mathbf{r}-\mathbf{r}'|}\,\mathrm d^3r'.
\label{eq.exx3}
\end{equation}

\subsection{The semi-local correlation contribution}

The self-interaction-free semi-local correlation energy contribution $E_c^{sl} [ \{ \pphi_{i \s} \} ]$ is defined by
\begin{equation} \label{coco1}
 E^{sl}_c[ \{ \pphi_{i \s} \} ] = \int g\rrp Q\rrp \, \mathrm d^3r',
\end{equation}
where
\begin{equation} \label{f2}
g\rr = 1 - \frac{\tau_W\rr}{\tau\rr}\zeta^2\rr 
\end{equation}
and 
\begin{equation}
Q\rr = n\rr e_c^\LSDA\rr. 
\end{equation}

For completeness, we list out all the quantities that are required to construct the function $g\rr$:\\
\begin{compactenum}[a)]
   \item kinetic energy density
\begin{equation}
\tau\rr=\frac{1}{2}\sum_{\substack{
i=1\\
\sigma=\up,\dn}}^{N_\sigma}|\nabla\varphi_{i\sigma}\rr|^2 ;
\end{equation}
 \item Von Weizs\"{a}cker kinetic energy density
\begin{equation}
\tau_W\rr=\frac{|\nabla n\rr|^2}{8 n\rr}=\frac{1}{2} |\nabla n^{\frac{1}{2}}\rr|^2 ;
\end{equation}
 \item spin polarization
$$ \zeta\rr=\frac{n_\up\rr-n_\dn\rr}{n_\up\rr+n_\dn\rr}$$.
\end{compactenum}
Taking the functional derivative based on Eq.~(\ref{coco1}) results in two parts:
\begin{eqnarray}
 \varphi^*_{i\sigma}\rr u^{sl}_{i\sigma}\rr
&=& \int \left(\frac{\delta g\rrp}{\delta\varphi_{i\sigma}\rr
}\right)Q\rrp\,\mathrm d^3r'\notag \\
& & +  \int g\rrp\left(\frac{\delta Q\rrp}{\delta\varphi_{i\sigma}\rr
}\right)\mathrm d^3r' \label{twopartscoco}
\end{eqnarray}
By denoting the constituent functions of the function $g\rr$ by
$\psi_1\rr = \tau\rr$, $\psi_2\rr = \tau_W\rr$ and $\psi_3\rr = \zeta\rr$, chain rule arguments lead to the
following expression:
 \begin{eqnarray}\label{psisum}
\left(\frac{\delta g\rrp}{\delta\varphi_{i\sigma}\rr }\right) = 
\sum_{l=1}^{3}\frac{\delta \psi_l\rrp}{\delta\varphi_{i\sigma}\rr } \frac{\delta g\rrp}{\delta \psi_l\rrp } \end{eqnarray}
Here we explicitly took into account the fact that $g\rr$ depends on $\psi_l\rr$ locally.

In order to obtain an analytical expression for $\delta g\rrp / \delta \pphi_{i\s} \rr$, which can then be inserted into Eq.~(\ref{twopartscoco}), one has to consider the three constituent functions $\psi_l\rr$ separately:\\\\
$l=1:$
\begin{align}
\frac{\delta\tau\rrp}{\delta\varphi_{i\sigma}\rr} &=
- \frac{1}{2} \delta(\mathbf{r}-\mathbf{r}') \lp[ \nabla'^2 \varphi^*_{i\sigma}\rrp + (\nabla' \varphi^*_{i\sigma}\rrp ) \cdot \nabla' \rp] \label{tau1}
\\
\frac{\delta g\rrp}{\delta \tau\rrp} &= \frac{\tau_W\rrp\zeta^2\rrp}{\tau^2\rrp}  \label{tau2}
\end{align} 
$l=2:$
\begin{align}
\nonumber \frac{\delta\tau_W\rrp}{\delta\varphi_{i\sigma}\rr} &= 
- \frac{\varphi^*_{i\sigma}\rrp}{2n^{\frac{1}{2}}\rrp} & \!\!\! \delta(\mathbf{r}-\mathbf{r}') \lp( \nabla'^2 (n^
{\frac{1}{2}}\rrp) + \right. \\
 & &\left. ( \nabla' n^{\frac{1}{2}}\rrp ) \cdot \nabla' \rp) \label{tauw1} 
\\
\frac{\delta g\rrp}{\delta \tau_W\rrp} &= - \frac{\zeta^2\rrp}{\tau\rrp}  \label{tauw2}
\end{align}
$l=3:$
 \begin{eqnarray}
\frac{\delta \zeta\rrp}{\delta\varphi_{i\sigma}\rr} &=&
\varphi^*_{i\sigma}\rrp\delta(\mathbf{r}-\mathbf{r}')\lp(\frac{\delta_{\sigma}-\zeta\rrp}{n\rrp}\rp)
\label{zeta1}\\ 
  \frac{\delta g\rrp}{\delta \zeta\rrp} &=& -\frac{2\tau_W\rrp\zeta\rrp}{\tau\rrp} \label{zeta2}
\end{eqnarray}
Here the operator $\nabla'$ denotes a gradient relative to the coordinate $\rrp$ and the quantity $\delta_{\sigma}$ distinguishes between the two spin channels by
\begin{equation}
 \delta_{\sigma}=\begin{cases}

  \,\,\,\, 1 & \text{if }\sigma=\uparrow\\
  -1 & \text{if }\sigma=\downarrow.
\end{cases}
\end{equation}
It is noted that the functional derivatives above have the presented form with respect to the \emph{occupied} orbitals only; derivatives with respect to unoccupied orbitals equal zero.

The derived relations  (\ref{tau1}) -  (\ref{zeta2}) now have to be inserted via Eq.~(\ref{psisum}) into Eq.~(\ref{twopartscoco}). 
By further employing a chain rule argument for the second term on the RHS of Eq.~(\ref{twopartscoco})
\begin{eqnarray}
 \frac{\delta Q\rrp}{\delta\varphi_{i\sigma}\rr}&=&\sum_{\tau=\up,\dn}\int
\frac{\delta Q\rrp}{\delta n_{\tau}(\mathbf{r}'')}\frac{\delta
n_{\tau}(\mathbf{r}'')}{\delta\varphi_{i\sigma}\rr}\,\mathrm d^3r'' \notag \\
& =&  \varphi^*_{i\sigma}\rr\frac{\delta Q\rrp}{\delta n_{\sigma}\rr}
=  \varphi^*_{i\sigma}\rr v_{c, \sigma}^\LSDA\rrp
\delta(\mathbf{r}-\mathbf{r}'), \notag\\
\label{secondpart}
\end{eqnarray}
one arrives at the final expression of the functional derivative of $E_c^{sl} [ \{ \pphi_{i \s} \} ]$:
\begin{align}
& \varphi^*_{i\sigma}\rr u^{sl}_{i\sigma}\rr  = & \notag\\
& -\frac{1}{2}\left[\left(\nabla^2\varphi^*_{i\sigma}\rr\right)\frac{\delta g\rr}{\delta \tau\rr}Q\rr
+\nabla\varphi^*_{i\sigma}\rr\cdot \nabla \left(\frac{\delta g\rr}{\delta \tau\rr}Q\rr  \right)\right] & \notag 
\\
& -\frac{\varphi^*_{i\sigma}\rr}{2n^{\frac{1}{2}}\rr} \left[\left(\nabla^2n^{\frac{1}{2}} \rr\right)
\frac{\delta g\rr}{\delta \tau_W\rr}Q\rr + \right. &\notag
\\ & \hspace{4.5cm} \left. \nabla n^{\frac{1}{2}}\rr\cdot \nabla \left(\frac{\delta g\rr}{\delta \tau_W\rr}Q\rr\right) \right]\notag
\\
& + \varphi^*_{i\sigma}\rr\
\left(\delta_{\sigma}-\zeta\rr\right)\frac{\delta g\rr}{\delta
\zeta\rr}e^\LSDA_c\rr  & \notag \\
& + \varphi^*_{i\sigma}\rr \,g\rr\ v_{c, \sigma}^\LSDA \rr &
\label{cctotal}
\end{align}
Equation~({\ref{cctotal}) corresponds to the functional derivative the way it was implemented into the
KLI/OEP-routine in the program package \darsec.

\subsection{The contribution $E^{iso}_{c} [ \{ \pphi_{i \s} \} ]$}
The correlation term $E^{iso}_{c} [ \{ \pphi_{i \s} \} ]$ is defined in Eq.~(\ref{eq.Eiso}). Let us denote
\begin{equation}
  P\rr= n\rr\lp(e_x^\LSDA\rr-e_x^{ex}\rr\rp)
\end{equation}
and recall that the LMF function $f\rr$ equals
\begin{equation} \label{f1}
f\rr = \frac{1-\frac{\tau_W\rr}{\tau\rr}\zeta^2\rr }{1+c\cdot t^2\rr} 
\end{equation}

In addition to the quantities $\tau$, $\tau_W$ and $\zeta$ introduced above, the function $f\rr$ additionally employs the so-called \emph{reduced density gradient}~\cite{Perdew1996}
\begin{eqnarray}
 t\rr&=& \left(\frac{\pi}{3}\right)^{\frac{1}{6}}\frac{a_0^{\frac{1}{2}}}{4 \Phi\rr} \frac{|\nabla
n\rr|}{n^{\frac{7}{6}}\rr} \notag \\
&:=& a\frac{t_n\rr}{\Phi\rr}
\end{eqnarray}
with
\begin{eqnarray}
 \Phi\rr=\frac{1}{2}\lp[\lp(1+\zeta\rr\rp)^{\frac{2}{3}}+\lp(1-\zeta\rr\rp)^{\frac{2}{3}}\rp]
\end{eqnarray}
and
\begin{eqnarray}
 a=\left(\frac{\pi}{3}\right)^{\frac{1}{6}}\frac{a_0^{\frac{1}{2}}}{4}=const.\notag
\end{eqnarray}
The exact relation
\begin{eqnarray}
 t_n^2\rr=\frac{8\tau_W\rr}{n^{\frac{4}{3}}\rr}
\end{eqnarray}
will be useful for later derivations.

Analogously to Eq.~(\ref{twopartscoco}), the application of the functional derivative with respect to the
KS-orbitals leads to two contributions:
\begin{eqnarray}
 \varphi^*_{i\sigma}\rr u^{iso}_{i\sigma}\rr
&=& \int \left(\frac{\delta f\rrp}{\delta\varphi_{i\sigma}\rr }\right) P\rrp\,\mathrm d^3r'\notag \\
& & +  \int f\rrp \left(\frac{\delta P\rrp}{\delta\varphi_{i\sigma}\rr }\right)\mathrm d^3r' \label{twopartsiso}
\end{eqnarray}
Moreover, an analogous relation to Eq. (\ref{psisum}) helps to rewrite the first part of this equation, only that now
one has to consider also the functions $\psi_4\rr = t_n^2\rr$ and $\psi_5\rr = \Phi\rr$:
\begin{eqnarray}
\left(\frac{\delta f\rrp}{\delta\varphi_{i\sigma}\rr}\right) =
  \sum_{l=1}^{5}\frac{\delta \psi_l\rrp}{\delta\varphi_{i\sigma}\rr
} \frac{\delta f\rrp}{\delta \psi_l\rrp \label{phisum}
} \end{eqnarray}
We evaluate each term separately and obtain:
\\$l=1:$
\begin{equation}\label{2tau2} 
\frac{\delta f\rrp}{\delta \tau\rrp} = \frac{\frac{\tau_W\rrp}{\tau^2\rrp}\zeta^2\rrp}{1+c\cdot t^2\rrp} 
= -\frac{\delta f\rrp}{\delta \tau_W\rrp} \frac{\tau_W\rrp}{\tau\rrp}
\end{equation}
$l=2:$
\begin{equation}
\frac{\delta f\rrp}{\delta \tau_W\rrp} = -\frac{\zeta^2\rrp}{\tau\rrp\lp(1+c\cdot t^2\rrp\rp)} \label{2tauw2}
\end{equation}
$l=3:$
\begin{equation} \label{2zeta2}
\frac{\delta f\rrp}{\delta \zeta\rrp} = -\frac{2\frac{\tau_W\rrp}{\tau\rrp}\zeta\rrp}{1+c\cdot t^2\rrp}  
\end{equation}
$l=4:$
\begin{align}
& \frac{\delta t_n^2\rrp}{\delta\varphi_{i\sigma}\rr} = 8 \frac{\delta\tau_W\rrp}{\delta\varphi_{i\sigma}\rr} \frac{1}{n^{\frac{4}{3}}\rrp}- \frac{32}{3} \frac{\tau_W\rrp\varphi^*_{i\sigma}\rrp}{n^{\frac{7}{3}}\rrp} \delta(\mathbf{r}-\mathbf{r}') \label{2tn1} \\
& \frac{\delta f\rrp}{\delta t_n^2\rrp} = - \frac{ca^2 f\rrp}{\Phi^2\rrp\lp(1+c\cdot t^2\rrp\rp)} \label{2tn2}
\end{align}
$l=5:$
 \begin{eqnarray}
  \frac{\delta \Phi\rrp}{\delta\varphi_{i\sigma}\rr} &=&\frac{\delta
\Phi\rrp}{\delta\zeta\rrp}\frac{\delta \zeta\rrp}{\delta\varphi_{i\sigma}\rr} \notag\\
 &=& \frac{1}{3}\lp[\lp(1+\zeta\rrp\rp)^{-\frac{1}{3}}-\lp(1-\zeta\rrp\rp)^{-\frac{1}{3}}\rp]\cdot \notag
\\
& &\varphi^*_{i\sigma}\rrp\delta(\mathbf{r}-\mathbf{r}')\lp(\frac{\delta_{\sigma}-\zeta\rrp}{n\rrp} \rp)
\end{eqnarray}
To avoid numerical instability due to the negative powers of $-\frac{1}{3}$, we multiply the above relation by $(1+\zeta\rrp)^{\frac{1}{3}}(1-\zeta\rrp)^{\frac{1}{3}}$ and then divide by the same term expressed in terms of the spin-densities. We
then obtain
\begin{align}
& \frac{\delta \Phi\rrp}{\delta\varphi_{i\sigma}\rr} = -\frac{1}{3}\lp[\lp(1+\zeta\rrp\rp)^{\frac{1}{3}}-\lp(1-\zeta\rrp\rp)^{\frac{1}{3}}\rp]\cdot \notag\\
& \hspace{1.2cm} \frac{n^{\frac{2}{3}}\rrp}{2^{\frac{2}{3}}\lp(n_\up\rrp
n_\dn\rrp\rp)^{\frac{1}{3}}}\varphi^*_{i\sigma}\rrp \cdot \delta(\mathbf{r}-\mathbf{r}')\lp(\frac{\delta_{\sigma}-\zeta\rrp}{n\rrp}\rp) \label{2phi1}
\\
& \frac{\delta f\rrp}{\delta \Phi\rrp} =
\frac{2ct^2\rrp f\rrp}{\Phi\rrp\lp(1+c\cdot t^2\rrp\rp)} \label{2phi2}
\end{align}

In order to compute the first term of Eq.~(\ref{twopartsiso}), one now has to evaluate all the contributions
originating from the different $\psi_l$ (Eqs.~(\ref{tau1}), (\ref{tauw1}), (\ref{zeta1}), (\ref{2tau2}), (\ref{2tauw2}), (\ref{2zeta2}), (\ref{2tn1}), (\ref{2tn2}), (\ref{2phi1}), (\ref{2phi2})) via the chain rule argument
(\ref{phisum}). 

Similar considerations are now used for the second term on the RHS Eq.~(\ref{twopartsiso}). By applying chain rule arguments only to the semi-local energy density part of $P\rr$, one arrives at the
following equation:
\begin{eqnarray}
 \frac{\delta P\rrp}{\delta\varphi_{i\sigma}\rr} &=& \sum_{\tau=\up,\dn}\int
\frac{\delta \lp(n\rrp e_x^\LSDA \rrp\rp)}{\delta n_{\tau}(\mathbf{r}'')}\frac{\delta
n_{\tau}(\mathbf{r}'')}{\delta\varphi_{i\sigma}\rr}\,\mathrm d^3r'' \notag \\
& &-\frac{\delta\lp(n\rrp e_x^{ex}\rrp\rp)}{\delta\varphi_{i\sigma}\rr}\label{chainruleP1}
\end{eqnarray}
While the first term contributes simply via the regular density-dependent LSDA exchange potential (similar to Eq.~(\ref{secondpart})), requires the second term explicit evaluation of the exact exchange energy density
\begin{equation}\label{exx_per_rho}
n\rrp e_x^{ex}\rrp \!\! = \!\! -\frac{1}{2} \!\! \sum_{\substack{
k,q=1\\
\upsilon=\up,\dn}}^{N_{\upsilon}} \!\! \int \!\!
\frac{\varphi_{k\upsilon}^*\rrp\varphi_{q\upsilon}\rrp\varphi_{k\upsilon} (\mathbf{r}'')\varphi_{
q\upsilon}^*(\mathbf{r}'')}{|\mathbf{r}'-\mathbf{r}''|}\mathrm d^3r''. 
\end{equation}
Therefore, Eq. (\ref{chainruleP1}) results in
\begin{eqnarray}
\frac{\delta P\rrp}{\delta\varphi_{i\sigma}\rr} & =& \varphi^*_{i\sigma}\rr v_{x,\sigma}^\LSDA\rrp
\delta(\mathbf{r}-\mathbf{r}') \notag \\
&+& \frac{1}{2}\sum_{k=1}^{N_{\sigma}}\delta(\mathbf{r}-\mathbf{r}')\varphi_{k\sigma}^*\rrp\int
\frac{\varphi_{k\sigma} (\mathbf{r}'')\varphi_{
i\sigma}^*(\mathbf{r}'')}{|\mathbf{r}'-\mathbf{r}''|}\,\mathrm d^3r'' \notag \\
&+& \frac{1}{2}\sum_{q=1}^{N_{\sigma}}
\frac{\varphi_{i\sigma}^*\rrp\varphi_{q\sigma}\rrp\varphi_{q\sigma}^*(\mathbf{r})}{|\mathbf{r}'-\mathbf
{r}|}.
\end{eqnarray}
Evaluating this expression with the corresponding integral in Eq.~(\ref{twopartsiso}) and adding the
previously derived first term, one arrives at the final expression for the functional derivative of
$E^{iso}_{c}[\{\varphi\}]$ with respect to the KS orbitals:
\begin{widetext}
\begin{align}
& \varphi^*_{i\sigma}\rr u^{iso}_{i\sigma}\rr = - \frac{f\rr}{2} \varphi^*_{i\sigma}\rr\,u^{exx}_{i\sigma}\rr 
 +\frac{1}{2}\sum_{j=1}^{N_{\sigma}} \varphi^*_{j\sigma}\rr\notag \int f\rrp\frac{\varphi_{i\sigma}^*\rrp\varphi_{j\sigma}\rrp}{|\mathbf{r}-\mathbf{r}'|} \,\mathrm d^3r' \notag 
 + \varphi^*_{i\sigma}\rr f\rr  v_{x,\sigma}^\LSDA \rr
\\
&  -\frac{1}{2} \left[\left(\nabla^2\varphi^*_{i\sigma}\rr\right)\frac{\delta f\rr}{\delta \tau\rr}P\rr +\nabla\varphi^*_{i\sigma}\rr\cdot \nabla \left(\frac{\delta f\rr}{\delta \tau\rr}P\rr  \right)\right] 
-\frac{\varphi^*_{i\sigma}\rr}{2n^{\frac{1}{2}}\rr}\left[\left(\nabla^2n^{\frac{1}{2}} \rr\right)\frac{\delta f\rr}{\delta \tau_W\rr}P\rr + \right. \notag
\\ 
& \hspace{12cm} \left. \nabla n^{\frac{1}{2}}\rr \cdot \nabla \left(\frac{\delta f\rr}{\delta \tau_W\rr}P\rr\right) \right]  \notag
\\
& -\frac{1}{3}\lp[\lp(1+\zeta\rr\rp)^{\frac{1}{3}}-\lp(1-\zeta\rr\rp)^{\frac{1}{3}}\rp]
\frac{\varphi^*_{i\sigma}\rr n^{\frac{2}{3}}\rr}{2^{\frac{2}{3}}\lp(n_\up\rr n_\dn\rr\rp)^{\frac{1}{3}}}
\lp({\delta_{\sigma}-\zeta\rr}\rp)\frac{\delta f\rr}{\delta \Phi\rr}\lp(e^\LSDA_x\rr-e^{ex}_x\rr\rp) \notag 
\\
& - \frac{2 \varphi^*_{i\sigma}\rr}{n^{\frac{4}{3}}\rr} \lp[ \nabla^2n\rr \lp(e^\LSDA_x\rr-e^{ex}_x\rr\rp)\frac{\delta f\rr}{\delta t_n^2\rr} -\frac{28}{3}\tau_W\rr \lp(e^\LSDA_x\rr-e^{ex}_x\rr\rp)\frac{\delta
f\rr}{\delta t_n^2\rr} + \right. \notag 
\\ 
& \hspace{10cm} \left. \nabla n\rr \cdot \nabla \lp(\lp(e^\LSDA_x\rr-e^{ex}_x\rr\rp)\frac{\delta f\rr}{\delta t_n^2\rr}\rp) \rp]  \notag
\\
& + \varphi^*_{i\sigma}\rr \cdot \frac{2ct^2\rr}{1+ct^2\rr}f\rr\lp(e^\LSDA_x\rr-e^{ex}_x\rr\rp) 
+\varphi^*_{i\sigma}\rr \lp({\delta_{\sigma}-\zeta\rr}\rp)\frac{\delta f\rr}{\delta \zeta\rr}\lp(e^\LSDA_x\rr-e^{ex}_x\rr\rp)
\label{eq.uisocomplete}
\end{align}
\end{widetext}

Finally, we 
note that when numerically implementing such complex expressions, questions of numerical stability may emerge. We found that implementing the von Weizs\"acker kinetic energy density as $\tau_W\rr = \frac{1}{2} |\nabla n^{\frac{1}{2}}\rr|^2$ and the quantity $\frac{\delta \Phi\rrp}{\delta\varphi_{i\sigma}\rr}$ as in Eq.~(\ref{2phi1}) is 
highly advantageous. In addition, we store $\tau_W\rr / \tau\rr$ as a separate quantity, enforcing the exact condition that it is never larger than 1. We also store separately the quantity $(1 + c t^2\rr)^{-1}$ and express $ct^2\rr / (1 + c t^2\rr)$ in terms of the former, to avoid the divergence of $t\rr$ at large distances.

\section{OEP/KLI comparison}\label{sec.app.oep_kli_comp}
This Appendix reports detailed numerical results for the total energies, $E_{tot}$, as well as the eigenvalues of the highest occupied KS state, $\eps_{ho}$, using the proposed local hybrid functional for selected systems:
the BH, Li$_2$, NH, and the N$_2$ molecules, as well as the C atom.
A multiplicative, local KS potential was obtained by employing the functional derivative of Eq.~(\ref{eq.uisocomplete}) either in the full OEP scheme or by the KLI approximation.
Table~\ref{table.app.oepcomp1} lists the absolute values of $E_{tot}$ for KLI and OEP, as well as the differences between results obtained with both schemes. Table~\ref{table.app.oepcomp2} provides the same comparsion for $\eps_{ho}$.

Note that the systems BH, Li$_2$, and N$_2$ are 
spin-unpolarized. 
Therefore, for $c=0$ the functional reduces to the LSDA xc functional (cf.\ Eq.~(\ref{eq.ec_ISO})) and thus no difference between KLI and OEP should occur. This is indeed the case, within numerical accuracy.

\begin{table}[h]
 \caption{Comparison of total energy, $E_{tot}$, using the suggested local hybrid functional in both the KLI and OEP schemes, as a function of $c$. All values are in Hartree.}\label{table.app.oepcomp1}

\begin{center}
\begin{tabular*}{\columnwidth}{@{\extracolsep{\stretch{1}}}*{5}{lrrrc}@{}}\hline
system 	& $c$ 	& KLI 		& OEP		& $E^{KLI}_{tot}-E^{OEP}_{tot}$	\\ \hline
C 	& 0 	& -37.4804 	& -37.4804 	& 0.0000 			\\
	& 0.5 	& -37.8108	& -37.8110	& 0.0002			\\
	& 2.5 	& -37.9494	& -37.9497	& 0.0003 			\\
BH 	& 0 	& -24.9768 	& -24.9768	& 0.0000			\\
	& 0.5 	& -25.2612 	& -25.2614	& 0.0002			\\
	& 2.5 	& -25.3983 	& -25.3988	& 0.0005			\\
Li$_2$  & 0 	& -14.7244 	& -14.7244	& 0.0000			\\
	& 0.5 	& -14.9809 	& -14.9810	& 0.0001			\\
	& 2.5 	& -15.1245 	& -15.1247	& 0.0002			\\
NH 	& 0 	& -54.7769	& -54.7770	& 0.0001			\\
	& 0.5 	& -55.1769 	& -55.1774	& 0.0005			\\
	& 2.5 	& -55.3555 	& -55.3563	& 0.0008			\\
N$_2$ 	& 0 	&-108.6958  	&-108.6958 	& 0.0000			\\
	& 0.5 	&-109.4464 	&-109.4474 	& 0.0009			\\
	& 2.5 	&-109.7593 	&-109.7609	& 0.0017
\end{tabular*}
\end{center}
 \end{table}
 \begin{table}[h]
 \caption{Comparison of highest occupied orbital energy $\eps_{ho}$ using the suggested local hybrid functional in both the KLI and OEP schemes, as a function of $c$. All values are in Hartree.}\label{table.app.oepcomp2}

\begin{center}
\begin{tabular*}{\columnwidth}{@{\extracolsep{\stretch{1}}}*{5}{lrrrc}@{}}\hline
system	& $c$	& KLI 		& OEP		& $\eps^{KLI}_{ho}-\eps^{OEP}_{ho}$	\\ \hline
C	& 0	& -0.2740	& -0.2736	&-0.0005					\\
	& 0.5	& -0.3067	& -0.3068	& 0.0001					\\
	& 2.5	& -0.3688	& -0.3691	& 0.0003						\\
BH 	& 0	& -0.2031	& -0.2031	& 0.0000					\\
 	& 0.5	& -0.2412	& -0.2415	& 0.0003					\\
 	& 2.5	& -0.3043	& -0.3047	& 0.0004					\\
Li$_2$ 	& 0	& -0.1189	& -0.1189	& 0.0000					\\
 	& 0.5	& -0.1286	& -0.1289	& 0.0002					\\
	& 2.5	& -0.1522	& -0.1527	& 0.0005					\\
NH 	& 0 	& -0.3157	& -0.3164	& 0.0007					\\
	& 0.5 	& -0.3770 	& -0.3783	& 0.0013					\\
	& 2.5 	& -0.4581 	& -0.4607	& 0.0025					\\
N$_2$ 	& 0	& -0.3825	& -0.3825	& 0.0000					\\
	& 0.5	& -0.4456	& -0.4447	&-0.0010					\\
	& 2.5	& -0.5463	& -0.5444	&-0.0019
\end{tabular*}
\end{center}
 \end{table}

\newpage

\bibliography{bibliography_ISO,skrefs}

\end{document}